\documentstyle[editedvolume,numreferences]{crckapb}

\def\simpropto{\hbox{$\, \buildrel {\scriptstyle \sim}\over 
{\scriptstyle \propto}\,$}}
\newcommand\simgreater{\buildrel > \over \sim}
\newcommand\simless{\buildrel < \over \sim}
\newcommand{\be}{\begin{equation}}
\newcommand{\ee}{\end{equation}}
\newcommand{\Msol}{\mbox{$M_{\odot}\;$}}

\begin{opening}
\title{THERMAL EVOLUTION OF ISOLATED NEUTRON STARS}
\author{Dany Page}
\institute{Instituto de Astronom\'{\i}a, UNAM, M\'exico City \\
           page@astroscu.unam.mx}
\end{opening}
\runningtitle{THERMAL EVOLUTION OF ISOLATED NEUTRON STARS}

\begin{document}

\begin{abstract}
\hspace{5pt} 
I give a general overview of the theory of neutron star cooling,
emphasizing the intuitive understanding of the effects of the various
physical ingredients, including the recently proposed accreted
envelopes and neutrino emission from Cooper pair breaking and
formation.
I describe how the present data may be compatible with both the `standard
model' and the fast cooling scenarios, with or without invoking extensive
presence of baryon pairing.
\end{abstract}

\section{Introduction}
        \label{sec:intro}

Understanding the interior of neutron stars is a challenge to human
intelligence and the study of their thermal evolution is one of the few
possible methods through which such understanding (or misunderstanding)
can be confronted with observations. 
The recent flow of data from {\em ROSAT} and the expected future avalanche of
high quality data from {\em XMM} and {\em AXAF}, among others, has given a 
strong impetus to the field, which has resulted in many new important
developments.
My purpose here is to present the basic picture, 
emphasizing an intuitive understanding of the
effect of the most important physical ingredients involved in this study.
I have tried to refer to most of the recent works in the field in order to
provide the reader with key entries into the literature.

\section{Basic Physics of Neutron Star Cooling}
        \label{sec:basics}

The thermal evolution of a neutron star is basically
given by the energy conservation equation
\be
\frac{dE_{th}}{dt} = -L
\ee
where $E_{th}$ is the thermal energy content of the star and $L$
the total luminosity, supplemented by the heat transport equation
(and general relativistic corrections) \cite{T86}.
$L$ naturally divides into the surface photon emission 
$L_{\gamma} = 4 \pi R^2 \sigma T_e^4$
and the internal neutrino emission $L_{\nu}$.
Internal heating mechanisms can be included in $L$ as
negative energy sinks $H$:
\be
L = L_{\gamma} + L_{\nu} - H.
\ee
The following subsections give a brief description of the most important
physics ingredients needed, whose recipes are put into an appropriate
stellar evolution code \cite{PB90,PA92} which is easily crunched by a 
workstation.

\subsection{The Equation of State}
           \label{sec:EOS}

The internal structure of the neutron star is determined by the
equation of state (EOS) which gives us two different inputs:
the radial density profile and the `chemical' composition, i.e.,
the type of particles present.
I will use the classical Friedman \& Pandharipande (FP) EOS \cite{FP81}
to determine the density profile and will add the core structure `by hand':
in principle, this is not the best thing to do, but there is so much uncertainty
regarding the other ingredients described below that using EOSs appropriate to
each scenario would not make much difference for this general description
of neutron star cooling, and the difference could always be canceled by
slightly adjusting another ingredient.

\subsection{The envelope}
           \label{sec:envelope}

Traditionally considered to be the layer beneath the atmosphere down to a
boundary density of $\rho_b = 10^{10}$ gm cm$^{-3}$, the envelope
is a region of utmost importance. 
Its temperature gradient is very large and
when the interior is isothermal it is the only region where a temperature
gradient still persists. 
The envelope has generally been considered to be formed of
iron-like nuclei and in this case, as a rule of thumb, the temperature at
its base $T_b$ and the effective temperature $T_e$ are related by \cite{GPE83}
\be
T_e \; \simpropto \; T_b^{0.5} 
\;\;\;\;\;\;\;\;\;\;\;\;\;\;\; 
T_e \approx 10^6 \; {\rm K} \; \longleftrightarrow \; T_b \approx 10^8 \; 
{\rm K.}
\ee
When the star is isothermal $T_b$ is equal to the interior temperature.

Recently, Chabrier, Potekhin \& Yakovlev \cite{CPY97} (see also
\cite{PCY97,SPYZ97}) showed that the presence of light elements
in the envelope strongly affects the heat transport
and results, for a given $T_b$ and at not too low a temperature, in
higher surface temperatures compared to the previous models with
beta-equilibrium matter.
{\em This strongly affects the cooling and can change radically 
the conclusions drawn by comparing models to the data}.

The pulsar magnetic field, if higher than about 10$^{11}$ G, also seriously 
affects
the envelope, but, when realistic surface field configurations 
are considered \cite{PS96,SY96}, the overall effect is not as large as has
sometimes been claimed, at least in the case of iron envelopes; the case of
magnetized accreted envelopes remains to be studied.

\subsection{Neutrino emission}
            \label{sec:neutrino}

The dominant neutrino emission processes occur in the core.
If neutrons and protons are the only baryons present and the proton
fraction is not too high, we are within the `standard model'
with the modified Urca process and its associated bremstrahlung brothers
giving a neutrino emissivity
\be
\epsilon_{\nu}^{MU} \; \cong \; 10^{20 - 21} \cdot \, T_9^8 \;\; 
                          {\rm erg \; cm^{-3} \; s^{-1}}.
\label{eq:slow}
\ee
Any change to this standard picture will almost always increase the
neutrino emission by orders of magnitude: a larger proton fraction
(which makes the direct Urca process possible \cite{LPPH91}), a pion or
kaon condensate, hyperons, quarks, etc... (see \cite{Pe92,Pr94}
for reviews).
It is thus convenient to divide the many possible scenarios into {\em slow} and 
{\em fast} neutrino cooling scenarios \cite{P94}, the former being the `standard
model' and the latter being any of the other ones
\footnote{I like to distinguish {\em scenario}
from {\em model}, the latter being a particular realization of the former
which the computer can crank with all its details. Nevertheless, I will still
use the term the `standard model' instead of `standard scenario'.}.
I will summarize the emissivity of the fast neutrino emission processes as
\be
\epsilon_{\nu}^{\cal N} = 10^{\cal N} \cdot \, T_9^6 \;\; 
                          {\rm erg \; cm^{-3} \; s^{-1}}
\label{eq:fast}
\ee
with $\cal N$ ranging from about 24 (kaon condensate) up to 27 (direct Urca)
\cite{P94} and a $T^6$ dependence instead of $T^8$ as in Eq.~\ref{eq:slow}.

Besides the controversy about which - if any - fast neutrino emission
process is allowed, there is still considerable uncertainty on
the modified Urca rate.
Voskresensky \& Senatorov \cite{VS86} proposed that medium effects may increase
it by up to 2 -- 3 orders of magnitude, a possibility which obviously has
a strong impact on the cooling \cite{SVSWW97}.

Although proposed many years ago \cite{FRS76,VS87}, the neutrino emission
by neutron (and proton) Cooper pair breaking and formation (PBF) has
mysteriously been completely neglected until very recently \cite{SVSWW97}.
Its neutrino emissivity can be comparable to or even higher than the
modified Urca value but it only acts at $T < T_c$ and is rapidly
suppressed when $T \ll T_c$: 
the result is a pulse of neutrino emission
and enhanced cooling of the layer undergoing the pairing phase transition
until  $T \ll T_c$.

Neutrinos are also emitted in the crust, but their effect is practically
negligible except at the early stages ($\sim$ 100 yrs).

\begin{figure}
\vspace{9.0cm}
\includegraphics{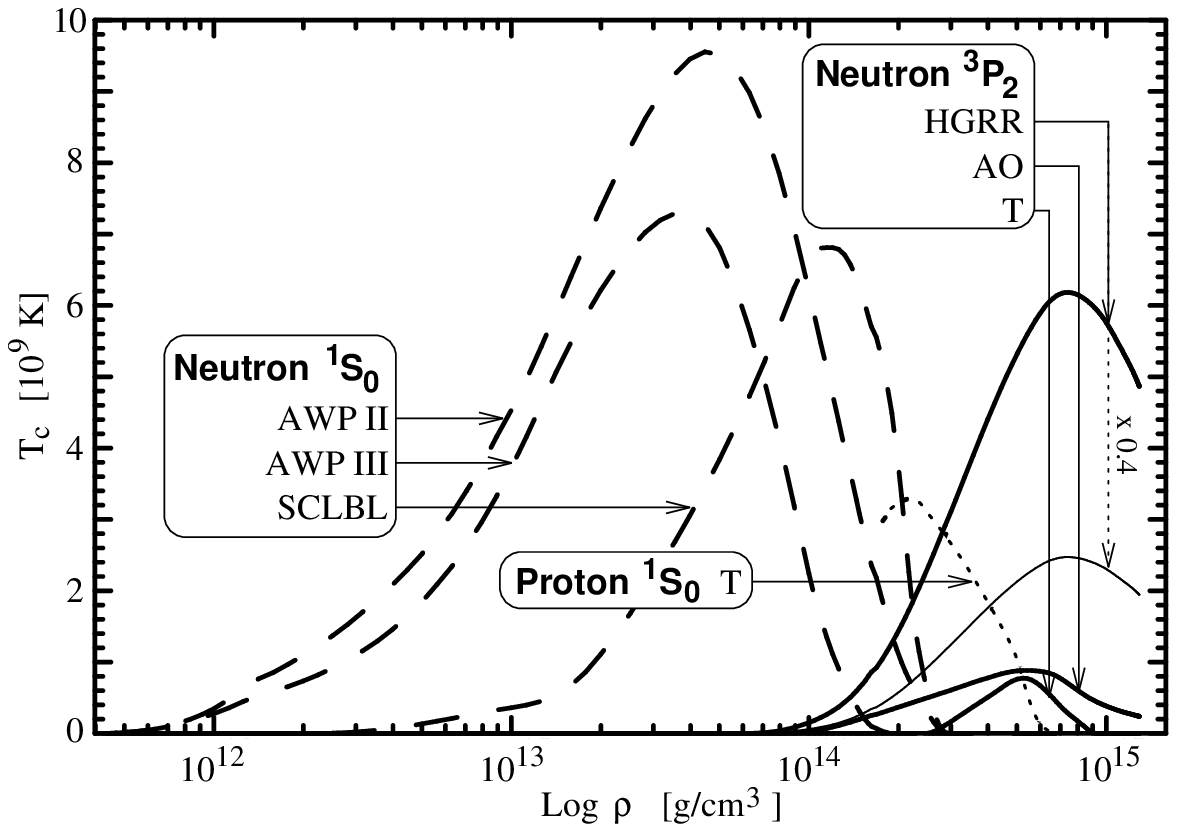}
\vspace{-0.3cm}
\caption{{\bf Pairing critical temperatures used in this paper.}
         \newline 
         Neutron $^1$S$_0$: AWP II \& AWP III from \protect\cite{AWP89},
                            and SCLBL from \protect\cite{SCLBL96} 
                            which are the most reliable values calculated 
                            to date.
         Neutron $^3$P$_2$: HGRR from \protect\cite{HGRR70},
                            and the same scaled down by 0.4,
                            AO from \protect\cite{AO85}, and
                            T from \protect\cite{T72};
                            the last two being more realistic than
                            the first one.
         Proton $^1$S$_0$:  T from \protect\cite{T73} which is close to the
                            better calculation of \protect\cite{WAP91}.
         The maximum density shown is the central density of my 1.4 
         \protect\Msol FP neutron star.
\label{fig:tc}}
\end{figure}

\subsection{Superfluidity and superconductivity}
            \label{sec:pairing}

In the pairing of the neutrons (= superfluidity) and protons 
(= superconductivity),
the Cooper pairs form in some angular momentum state which is usually
$^1$S$_0$ at low momentum and a mixed $^3$P$_2$ -- $^3$F$_2$ state at
higher momentum \cite{S97}.
As shown in Figure~\ref{fig:tc}, there are still large uncertainties
on the values of the critical temperatures $T_c$ at which these pairing phase 
transitions occur, particularly in the core.
The pairing phenomenon introduces a gap $\Delta$ in the single
particle excitation spectrum which, when appropriately defined,
is related to $T_c$ by $\Delta(0) \simeq 1.75 T_c$. 
At $T \ll T_c$ this gap produces a strong
suppression of the specific heat of the paired component and of the
neutrino emission processes in which this component participates:
\be
c_v^{(N)}
\; \rightarrow \; 
c_v^{(P)} = {\cal R}_c \cdot c_v^{(N)} 
\;\;\;\;\;\;\;\;\;\;\;\;\;\;\;\;\;\;\;
\epsilon_{\nu}^{(N)} 
\; \rightarrow \;
\epsilon_{\nu}^{(P)} = {\cal R}_{\nu} \cdot \epsilon_{\nu}^{(N)}
\label{eq:suppress}
\ee
where `N' stands for `normal' and `P' for `paired'.
The suppression coefficients ${\cal R}_c$ and ${\cal R}_{\nu}$, have
been described in detail by Levenfish \& Yakovlev \cite{LY94a,LY94b,YL95}.

\subsection{Internal Heating}
            \label{sec:heating}

Several mechanisms of internal heating have been proposed, for example:
friction due to the differential rotation of the crustal neutron $^1$S$_0$ 
superfluid \cite{AAPS84}; dissipative processes due to the core proton 
$^1$S$_0$ 
superconductor vortex lines \cite{SS93}; release of `chemical'
energy due to the readjustment of the chemical equilibrium of the core
 induced by the spin-down of the pulsar \cite{R95}.
The first of these is potentially the most efficient for young pulsars and has
already been studied in some detail \cite{SL89,USNT93,VRLE95}.
I will adopt it in a simple version \cite{P97}, writing the heating rate as
\be
H(t) = J_{44} \cdot 10^{40} \cdot 
       \left( \frac{t+\tau_0}{100 \; {\rm yrs}} \right)^{-3/2} \; \;
       \rm erg \; s^{-1}
\label{equ:heating}
\ee
where $t$ is the pulsar age, $\tau_0 = 300$ yrs a typical spin-down time scale,
and $J_{44}$ the differential angular momentum of the frictionally coupled
crustal neutron superfluid in units of 10$^{44}$ g cm$^2$ rad s$^{-1}$.
This expression assumes a standard spin-down rate from magnetic dipole radiation
and is similar to the one used in \cite{SL89,USNT93}.
This heating is distributed within the superfluid layers of the inner crust.
I adopt the value $J_{44} = 3.1$ which corresponds to moderately strong heating,
compatible with the size of the crust for the FP EOS.
Very different heating rates are of course possible \cite{VRLE95}.

\begin{figure}
\vspace{12.25cm}
\includegraphics{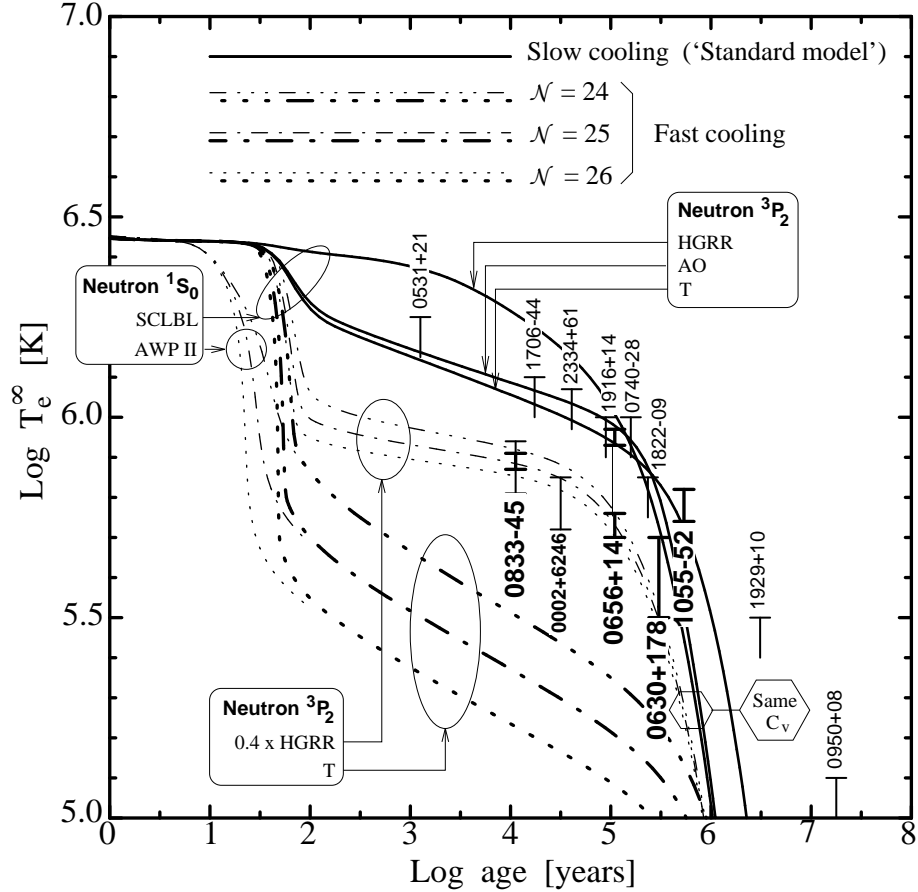}
\vspace{-0.2cm}
\caption{{\bf Typical behavior of slow (`standard') and fast cooling scenarios.}
\newline
1.4 \Msol neutron stars with the FP EOS.
The cases with $\cal N$ = 24, 25, and 26 (see Eq. \protect\ref{eq:fast})
correspond approximately to the effect of a kaon condensate, pion
condensate, and the direct Urca process (with hyperons or nucleons), 
respectively.
The various curves, within each scenario, show the effect of various
assumptions about pairing, following the notations of
Figure \protect\ref{fig:tc}:
all models use the proton $^1$S$_0$ $T_c$ `T', and the
neutron $^1$S$_0$ and $^3$P$_2$ $T_c$'s are as labeled.
All models have non-magnetized iron envelopes from \protect\cite{CPY97}.
Neither PBF neutrino emission nor heating are included.
\newline
The main effect of pairing in the crust (neutron $^1$S$_0$) is to shorten
the length of the early plateau.
Core pairing suppresses the neutrino emission, which results in a higher $T_e$
during the neutrino cooling era (age from $\sim$ 100 to $\sim$ 10$^5$ yrs),
and the specific heat, which results in faster cooling during the photon cooling
era (age above $\sim$ 10$^5$ yrs).
The reduction of the specific heat during the neutrino cooling era does not
show up as much as during the photon cooling era due to the small slope
of the curves at this phase.
\newline
The references to the data can be found in \protect\cite{P97}:
in short, the bigger the label the better the data.
All points 
are really upper limits (in several cases based on a non-detection of
the pulsar) but for the radio pulsars 0833-45 (Vela), 0656+14, 0630+178 
(Geminga) \protect\cite{KL97},
1055-52, and the neutron star 0002+6246, there is good evidence that the
observed X-rays are from surface thermal emission. Uncertainty on the
temperature estimate is illustrated in the case of PSR 0656+14 where
two values are reported.
For more details see \protect\cite{T97,P97}.
}
\label{fig:cool1}
\end{figure}

\section{An overview of the various scenarios}
         \label{sec:results}

\begin{figure}
\vspace{7cm}
\includegraphics{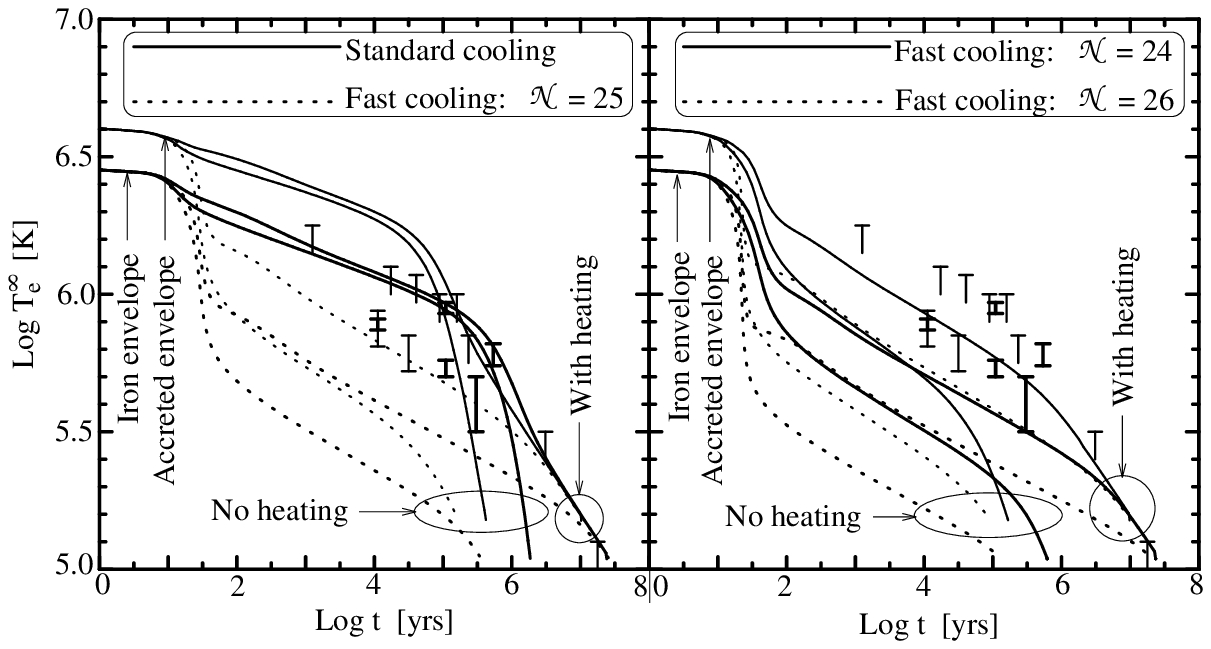}
\caption{{\bf Effect of an accreted envelope and heating on the cooling.}
\newline
1.4 \Msol neutron stars star with the FP EOS.
The fast neutrino emission is plugged in as described in 
\S~\protect\ref{sec:neutrino}.
All models use $T_c$'s from `AWP II' \protect\cite{AWP89} and `T' 
\protect\cite{T72}
for neutrons, and `T' \protect\cite{T73} for protons 
(see Figure~\protect\ref{fig:tc}): 
in these models there is {\em no} pairing in the inner core.
\newline
The presence of light elements in the envelope raises the surface temperature
during the neutrino cooling era and then hastens the cooling during the photon
cooling era. The internal heating works at all times but is of course
more efficient when the heat content, i.e., interior temperature,
of the star is low. When the initial heat content of the star has been
completely lost the cooling is independent of the previous history and envelope
structure since the luminosity is simply equal to the heating rate.
Notice that the temperature estimate of Geminga had been argued to require 
extensive pairing 
in the core to be compatible with cooling models \protect\cite{P94}, 
both `standard' and fast: 
in models with accreted envelopes, as shown here and in 
\protect\cite{CPY97,PCY97},
{\em extensive baryon pairing in the core is not necessary}.
The data points are labeled in Figure~\protect\ref{fig:cool1}.
}
\label{fig:cool2}
\end{figure}

\begin{figure}
\vspace{7cm}
\includegraphics{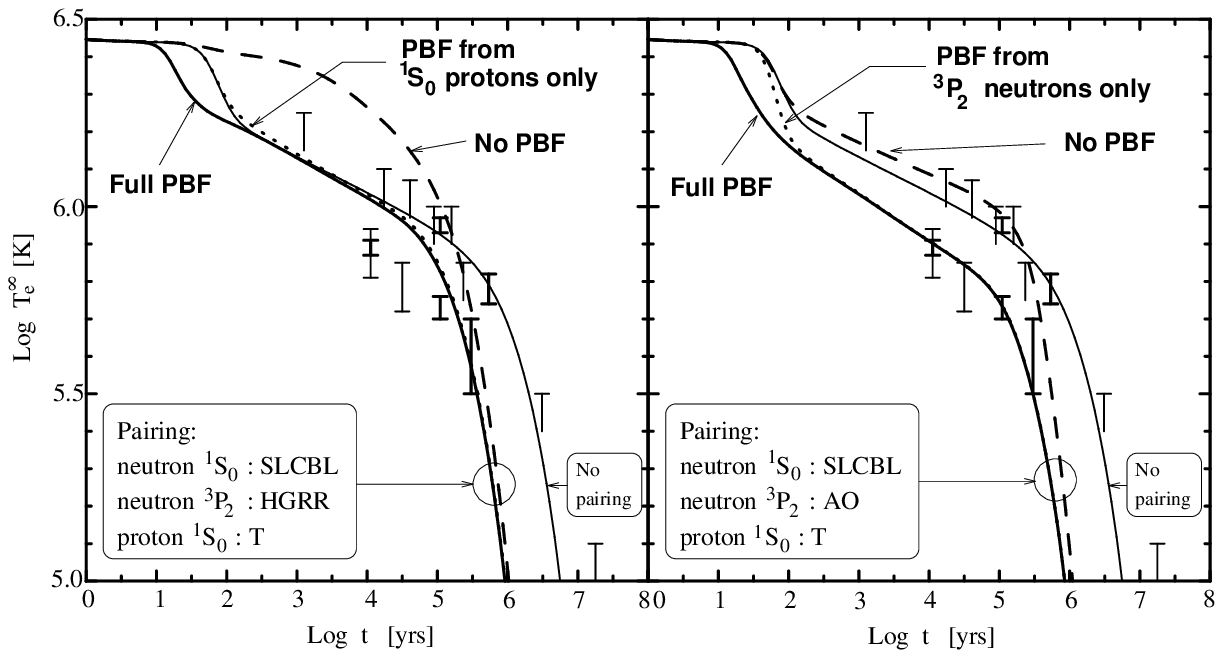}
\caption{{\bf Effect of the neutrino emission by the Pair Breaking and Formation
(PBF) process on the `standard' cooling.} 
\newline
1.4 \Msol neutron stars with the FP EOS.
The pairing $T_c$'s are as labeled (see Figure~\protect\ref{fig:tc}).
These models have an iron envelope and no heating is included.
\newline
The PBF emission in the crust hastens the isothermalization
but has little effect later on.
Once the star is isothermal, the cooling may be driven by the PBF emission from
either core proton $^1$S$_0$ pairing (left panel) or core neutron $^3$P$_2$ 
pairing
(right panel): it is the component with the lowest $T_c$ which dominates since
its PBF neutrino emission happens later. 
In these particular choices of pairing, the neutron $^3$P$_2$ PBF emission in 
the
right panel is more efficient than the proton $^1$S$_0$ PBF emission in the left
panel since a larger mass of paired matter is involved.
The data points are labeled in Figure~\protect\ref{fig:cool1}.
}
\label{fig:cool3}
\end{figure}

As the previous section made clear, we have a plethora of effects
which may affect the thermal evolution of a neutron star.
Several of them have been studied in detail in the literature while others
have been proposed, or simply unearthed, very recently and need more 
consideration.
I will show here a series of cooling curves which illustrate these various 
effects
in order to see them at work and try to understand them intuitively.
Figures~\ref{fig:cool1} to \ref{fig:cool3} should be deciphered by careful
comparison with Figure~\ref{fig:tc}.
Fast neutrino emission, when used, is plugged in at densities
$\rho > \rho_{cr} = 1.1 \cdot 10^{15}$ gm cm$^{-3}$.

\subsection{The general pre-1996 picture}
            \label{sec:pre-1996}

I show in Figure~\ref{fig:cool1} a set of curves to illustrate the main
cooling scenarios and the effect of baryon pairing.
At an early stage ($t \simless 10 - 100$ yrs) all models have the same
surface temperature: the surface does not know yet what
is happening in the core, because of the finite time scale for heat diffusion
through the crust \cite{NT87}, and its temperature is controlled by plasma
neutrino emission in the outer crust and by surface photon emission. 
All models use the same FP EOS and thus have the same crust and hence
exactly the same surface temperature at this stage.

After this early {\em plateau} the surface temperature decreases more or less 
strongly,
depending on the evolution of the core: 
this is the {\em isothermalization phase} during which the heat
content of the crust (and outer core) flows into the core,
where it is evaporated in neutrinos.
At the end of this phase the temperature profile inside the star is flat,
i.e., the star is isothermal, except for the strong temperature
gradient within the envelope.
The time needed for isothermalization is 
$\approx D^2 \cdot C_v/\lambda$
where $D$ is the thickness of the insulating layer (= crust + outer core),
$C_v$ its specific heat, and $\lambda$ its thermal conductivity.
Observation of a neutron star at this stage would directly `measure' the 
thickness
$D$ \cite{LVRPP94}.
However, this requires accurate pairing gaps in the crust
to calculate $C_v$ since pairing reduces it.
The two sets of fast cooling curves in
Figure~\ref{fig:cool1} with the neutron $^1$S$_0$ pairing from AWP II
and SCLBL show it clearly: in the AWP II case the gap extends to lower 
densities,
implying a lower specific heat and thus an earlier temperature drop.
Moreover, neutrino emission by the PBF process (see \S~\ref{sec:pbf}) 
complicates the situation.
In the case of a strange star \cite{G97}, this plateau lasts only about a year
\cite{P92} due to the absence of an inner crust, i.e., here $D$ is very small.

Once the star is isothermal we can distinguish a {\em neutrino cooling era}
followed by a {\em photon cooling era} depending on which energy loss
mechanism is driving the evolution: the change from the former to the latter 
shows
itself as a change in the slope of the cooling curves, occuring around
$t \sim 10^5$ yrs.

During the neutrino cooling era we see a clear difference between the
`standard model' and the fast cooling scenarios.
Within the fast cooling scenarios, the cases with the neutron $^3$P$_2$ $T_c$
from `T' \cite{T72} show the fast cooling unsuppressed: this gap, as well
as the proton gap, vanishes within the inner core where fast neutrino emission
occurs (Figure~\ref{fig:tc}).
The resulting surface temperatures are much below the observed ones and it had
been proposed \cite{PA92,PB90} that pairing gaps extending down to the center
of the star could control the fast neutrino emission and keep the star much 
warmer.
The suppression of the neutrino emission by the gap is of course very sensitive
to $T_c$, higher $T_c$'s resulting in higher surface temperatures 
\cite{PB90,PA92}.
One could actually use this to `measure' $T_c$ in the core of a neutron star by
fitting the cooling curves to the data \cite{GYS94,LY96,P95,PA92}.
This is what I have done in the models with neutron $^3$P$_2$ $T_c$ labeled
as `0.4 x HGRR': comparison with Figure~\ref{fig:tc} shows that $T_c$'s around
$2 \times 10^9$ K are needed.
This `measurement' needs exact suppression factors (Eq.~\ref{eq:suppress})
as used recently in \cite{GY93,GYS94,LY96,P95,P97} and implies values of $T_c$
higher than when simple Boltzmann factors are used as, recently, in 
\cite{PA92,PB90,SWWG96,SVSWW97,UNTMT94,UTN94}.
Notice that when the fast neutrino emission is suppressed by pairing,
the resulting cooling is only very weakly dependent on the exact cooling agent
(here the value of $\cal N$),
the EOS, and the critical density at which this agent starts operating 
\cite{P95}.
Within the `standard model' the effect of pairing is similar but less 
spectacular:
the curve with the neutron $^3$P$_2$ $T_c$ from `AO' \cite{AO85} is slightly 
warmer than the one with `T' \cite{T72} since the latter $T_c$ is lower and
also vanishes in the inner core, while the curve with the $T_c$ `HGRR'
\cite{HGRR70} is much warmer because the neutrino emission has been suppressed
very early on, in the whole core, due to the high value of $T_c$.

During the photon cooling era the temperature decreases much faster (on a log
scale) and the different cooling scenarios largely overlap: observations of
neutron stars of age $\simgreater 10^5$ yrs do not allow us to distinguish
between them \cite{P94}.
The cooling rate now depends crucially on the total specific heat
(and the structure of the envelope, as shown in \S~\ref{sec:accreted}).
For example, the two `standard' cooling curves `HGRR' and `AO' merge at this
stage, despite the very different $T_c$'s,
because in both cases the neutrons are paired in the whole core
and $T << T_c$, while the `T' curve cools more slowly since its inner core
neutrons are not paired and its specific heat is thus larger.
Moreover, notice that the fast cooling models with the `0.4 x HGRR' $T_c$
asymptotically approach the `standard' cooling curves `HGRR' and `AO':
at this time they all have the neutrons paired in the whole core and thus 
the same specific heat.

It is also most instructive to look at detailed temperature profiles for the
various scenarios: many have been published, for example in 
\cite{LVRPP94,MP96,NT87,USNT93,VR91} and the models of this paper
are available on the Web.

Magnetic fields in the envelope do alter the picture somewhat, but not
strongly \cite{PS96,SY96}: light elements in the envelope and
PBF neutrino emission have much more dramatic effects as shown in the 
next section.

\subsection{New developments}
            \label{sec:recent}

The picture presented in the previous section may change dramatically due
to three new ingredients introduced recently: the possible presence of light 
elements
in the upper layers of the neutron star, the neutrino emission by the 
breaking and formation of Cooper pairs, and the possibility of substantial
in-medium enhancement of the modified Urca process.

\subsubsection{Accreted envelopes (and heating)
               \label{sec:accreted}}

The light elements in the envelope increase the heat transport,
resulting in a higher $T_e$ for a given $T_b$, as long as 
$T_e$ is not too low \cite{CPY97,PCY97,SPYZ97}.
During the early stage and the neutrino cooling era the surface
temperature simply follows the evolution of the interior temperature
and does not affect the cooling rate.
As a result, at these stages a neutron star with an accreted envelope 
will undergo an evolution parallel to the evolution of the same neutron 
star with an iron envelope but with a higher $T_e$.
When photon cooling takes over the situation is reversed, since
for a given $T_b$ (i.e., interior temperature and thus
specific heat) the surface emission is stronger in the case of an 
accreted envelope because $T_e$ is higher.
Thus, the cooling trajectories of these two neutron stars cross and
the temperature of the one with an accreted envelope drops much faster.
This behavior is illustrated in Figure~\ref{fig:cool2} above and, 
even more clearly, in the Figure 2 of \cite{SPYZ97}.

If, moreover, internal heating is included, one can obtain
higher temperatures at all ages (nothing new in this of course,
see \cite{SL89,USNT93,VRLE95}).
This is extremely important for the fast cooling scenarios for which
the results of Figure~\ref{fig:cool2} show that {\em it is possible to obtain 
temperatures in agreement with the presently available data with
models where the gap vanishes within the inner core} \cite{P97},
in contradistinction to the results of \S~\ref{sec:pre-1996}.
The case $\cal N$ = 24 (i.e. a kaon condensate or a pion condensate with
its neutrino emission reduced by medium effects \cite{UNTMT94}) appears to be 
the most favorable but even the direct Urca process
($\cal N$ = 26 - 27) could be acceptable.

\subsubsection{Neutrino emission from Cooper pair breaking and formation
               \label{sec:pbf}}

Besides its suppressive effect, the occurrence of pairing does induce a strong, 
but short lived, neutrino emission \cite{FRS76,VS87} due to the constant 
breaking
and formation of Cooper pairs (\S~\ref{sec:neutrino}).
I show in Figure~\ref{fig:cool3} `standard' cooling curves with these PBF
mechanisms included: the results are quite sensitive to the values of $T_c$
for neutrons and/or protons in the core but the effect can be sufficient to
lower the star temperature significantly.
One sees that, {\em within the `standard' cooling scenario with the 
(also `standard') PBF
neutrino emission taken into account there may be no need of any `exotic'
process} (see Schaab {\em et al.} \cite{SVSWW97}).
Notice, however, that none of the models shown in Figure~\ref{fig:cool3} 
include any heating or an accreted envelope: inclusion of these
could spoil the apparent agreement with the data.

\subsubsection{Enhanced modified Urca process 
               \label{sec:enhancedMUrca}}

If, as Voskresensky \& Senatorov \cite{VS86} proposed, the modified
Urca rate is much more efficient than usually assumed, then all `standard'
cooling curves are pushed down and {\em all data, as presently interpreted, 
could
be compatible with the `standard model'} 
(see Schaab {\em et al.} \cite{SVSWW97}).
The actual efficiency of the modified Urca process is a very delicate, 
and controversial, problem but if neutron (or proton) pairing is present 
within the whole core it would be suppressed anyway and possibly
still result in too high temperatures.

\section{Conclusions}
         \label{sec:conclusion}

Apparently, our misunderstanding of the neutron star interior has progressed
as much as our understanding.
The present data seem to be compatible with both the standard scenario and the
fast cooling scenarios, depending on the other assumptions made in 
building the models.
However, the situation is far from desperate:

\smallskip
\noindent
-- On the observational side, future - and multiwavelength - observations will
certainly be able to determine the chemical composition of the atmosphere
\cite{PZTN96}, and thus probably of the envelope, of cooling neutron stars.
Observations of old pulsars will refute or confirm the presence of
internal heating \cite{PSC96} and help us to pin down its nature \cite{VRLE95}.

\smallskip
\noindent
-- On the theoretical side, what is badly needed, but yet can reasonably be
expected to be obtained, is a modified Urca rate that everybody will agree upon 
(at least the order of magnitude) and reliable
calculations of the pairing gaps for neutrons and protons in `standard' matter.
This will allow us to make more definite predictions for the standard scenario.
Moreover, much work is constantly being done on all the fast cooling scenarios.

Finally, the good news is that, thanks to the accreted envelopes, the fast 
cooling 
scenarios are viable without having to invoke the doubtful presence of
pairing at the extreme densities reached in `exotic' neutron stars \cite{P97}.

\bigskip
\noindent
Data files for all cooling curves shown here are available on the Web at:
\newline
\begin{small}
{\bf http://www.astroscu.unam.mx/neutrones/NS-Cooler/NS-Cooler.html}
\end{small}

\bigskip
\noindent
{\bf Acknowledgments.}
This work was supported by grants from UNAM-DGAPA (IN-105495)
and CONACYT (2127P-E9507).


\newpage
{\bf NOTE:} four references not yet published are available on the Web:
\begin{description}

\item[[9]]
Glendenning, N. 1997,
contribution to these proceedings
\newline
{\bf e-print: astro-ph/9706236}

\item[[30]]
Potekhin, A. Y., Chabrier, G., \& Yakovlev, D. G. 1997,
A\&A, in press 
\newline
{\bf e-print: astro-ph/9706148}

\item[[36]]
Schaab, Ch., Voskresensky, D., Sedrakian, A.D., Weber, F. \& Weigel, M. K. 1997,
A\&A, in press
\newline
{\bf e-print: astro-ph/9605188}

\item[[39]]
Shibanov, Yu. A., Potekhin, A.Y., Yakovlev, D.G., \& Zavlin, V.E. 1997,
contribution to these proceedings.
\newline
{\bf Preprint at: http://stella.ioffe.rssi.ru/NSG/NSG-Pub1.html}

\end{description}

\end {document}